\documentclass{epl}

\newcommand{\myscalebox}[1]{\scalebox{0.425}[0.425]{#1}}

\title{On random graphs and the statistical mechanics of granular matter}
\author{Johannes Berg\inst{1} and Anita Mehta\inst{2}}

\institute{
\inst{1}{Abdus Salam International Centre for Theoretical Physics, 
34100 Trieste, Italy}\\
\inst{2}{S N Bose National Centre for Basic Sciences, Block JD
Sector III, Salt Lake, Calcutta 700 098, India}
}

\pacs{05.20.-y}{Classical statistical mechanics}
\pacs{45.70.Cc}{Static sandpiles; granular compaction}
\pacs{75.10.Nr}{Spin-glass and other random models}

\begin{document}

\maketitle
 
\begin{abstract}
The dynamics of spins on a random graph with ferromagnetic three-spin 
interactions is used to model the compaction of granular matter under a 
series of taps. Taps are modelled as the random flipping of a small 
fraction of the spins followed by a quench at zero temperature. We find 
that the density approached during a logarithmically slow compaction 
 - the random-close-packing density - corresponds to a dynamical phase 
transition. We discuss the role of cascades of successive spin-flips in 
this model and link them with density-noise power fluctuations observed 
in recent experiments. 
\end{abstract}

Analogies with glasses and spin glasses have long \cite{first} 
been made to
explain the complex properties of granular media, but it was
not until the seminal experiments of the Chicago group \cite{nagel}
on granular compaction were carried out that serious
theoretical attempts were made to quantify such analogies. 
These fall into roughly two classes: lattice-based
models \cite{amgcb,tetris,paj}
in a finite-dimensional space (which in general do not admit analytic 
solutions) or mean field models \cite{jorge,cugliandolo} (where each 
site interacts with a large number of other sites). In this Letter 
we argue that models on random graphs may be used to describe 
those aspects of the behaviour of granular matter which 
depend on the {\em finite} connectivity of the (disordered) grains,
while still  
remaining analytically accessible. In particular this approach allows us 
to relate phenomena in the compaction of granular matter, such as 
the random-close-packing density, which is reached asymptotically after   
many taps, to the structure of the free-energy landscape of spin models 
on random graphs. 
  
The concept of geometric frustration is 
thought to be a crucial feature of the behaviour of granular matter. 
It refers to the existence of voids in the granulate, which cannot be 
filled in by neighbouring particles due to geometric constraints, 
either on the 
mobility of these particles, or on their compatibility in shape or size  
with the void. A class of models of this feature restrict the number of 
occupied neighbouring sites of a particle on a lattice, either dynamically  
\cite{kobandersen} or statically \cite{birolimezard}. 
More generally, and expressed in the language of interacting spins 
(e.g. spin up for the occupied 
site of a lattice gas model), geometrical frustration may be written in terms 
of ferromagnetic nearest-neighbour interactions and anti-ferromagnetic 
next-nearest 
neighbour interactions \cite{sethna}. Models on random graphs may 
be viewed as the Bethe-approximation to finite-dimensional models 
of this type. The common feature of these models when translated 
into spin-language is the existence of multi-spin 
interactions on plaquettes on a random graph \cite{aging}. 

In the following we consider the simplest possible case of a multi-spin model 
defined on the plaquettes of a random graph, a 3-spin Hamiltonian on a 
random graph where $N$ binary spins $S_i=\pm1$ interact in triplets 
\begin{equation}
\label{hdef}
H=-\rho N=-\sum_{i<j<k} C_{ijk} S_i S_j S_k 
\end{equation}
where the variable $C_{ijk}=1$ with $i<j<k$ denotes the presence 
of a plaquette connecting sites $i,j,k$ and $C_{ijk}=0$ 
denotes its absence.   
Choosing $C_{ijk}=1(0)$ randomly with probability 
$2c/N^2$ ($1-2c/N^2$) results in a random graph, where the number 
of plaquettes connected to a site is distributed with a 
Poisson distribution of average $c$. The connection with granular compaction
is made in accordance with Edwards' thermodynamic 
hypothesis \cite{sam}: We interpret the
local contribution to the energy in different 
configurations of the spins as the volume occupied by grains 
in different local orientations. 
In modelling terms, its random structure is 
an obvious advantage in the context of the disordered nature 
of granular media: Additionally, the locally fluctuating 
connectivity may be thought of as modelling the range of 
coordination numbers of the grains \cite{pre}. 

This Hamiltonian has recently been studied on a random graph 
in the context of satisfiability problems in combinatorial 
optimization \cite{hypersat}, and on a 2D triangular lattice 
\cite{newman}. It has a trivial ground state where all spins 
point up and all plaquettes are in the configuration 
$+++$ giving a contribution of $-1$ to the energy. 
Yet, \emph{locally}, plaquettes 
of the type $--+,-+-,+--$ (satisfied plaquettes) also give the same 
contribution, 
although one may not be able to cover the 
entire graph with these $4$ types of plaquettes in equal 
proportions. In this case, which occurs 
for $c>c_c \sim 2.75$ \cite{hypersat} most ground states will be 
ferromagnetic - corresponding to a state with long-range order and a 
crystalline state of the granular medium;
reaching the state 
$+++$ from $--+,-+-,+--$ requires two spin-flips and 
a crossing of an energetic barrier.

We argue that the $3$-spins ferromagnetic Hamiltonian is a suitable 
model-Hamiltonian for two reasons: 
The mechanism by which energy 
barrier has to be crossed in going from one metastable state to 
another aims to model the situation where a grain 
in order to reach a void in a granular assembly has to 
push apart temporarily other grains thus increasing the volume. This   
has recently been argued to be an important ingredient in models 
of granular compaction \cite{jpcm}. 

The feature of the model responsible for the slow dynamics, however, is 
the degeneracy of the four configurations of plaquettes with 
$s_i s_j s_k=1$ resulting in competition between satisfying plaquettes 
(dense arrangements of particles) \emph{locally}  
(all states with even parity may be used, resulting in a large 
entropy) and \emph{global} order (only the $+++$ state may be used). 
When grains are shaken, 
they rearrange locally, but locally dense configurations can be mutually
incompatible: voids appear between densely packed clusters
due to mutually incompatible grain orientations (geometric frustration) 
- leading to competition between locally dense packings and global 
crystalline order. 

To mimic the action of tapping leading to compaction 
we choose the following dynamics:  
Each tap consists of two phases. First, in the {\em dilation}
phase,
 a small fraction 
(taken to be $10^{-4}$) of the sites is chosen at random and 
have their spins flipped. This corresponds to the rapid acceleration 
of the sample during tapping. Then, the system undergoes several 
(in our case $3$) Monte-Carlo sweeps at zero temperature: Each sweep 
consists of choosing $N$ sites in series and flipping their spins with 
probability 1 if the flip lowers the energy, with probability $1/2$ 
if the flip leaves the energy unchanged. 
This phase corresponds to the {\em relaxation} process after the tap, 
where the particles reach a new locally stable configuration. 

This dynamics is a simplified version, suitable for spin models, of the 
tapping dynamics used in cooperative Monte Carlo simulations
of sphere shaking \cite{prl}.
It has been recently introduced independently 
in the context of spins models of granular compaction by Dean and Lef{\`e}vre 
\cite{dean} (however their $2$-spin models lack the competition 
between local and global order). In the context of combinatorial 
optimization it corresponds to the Walksat-algorithm, 
known to be one of the most efficient algorithms for 
the solution of satisfiability problems \cite{walksat}. 

\begin{figure}
\myscalebox{\onefigure{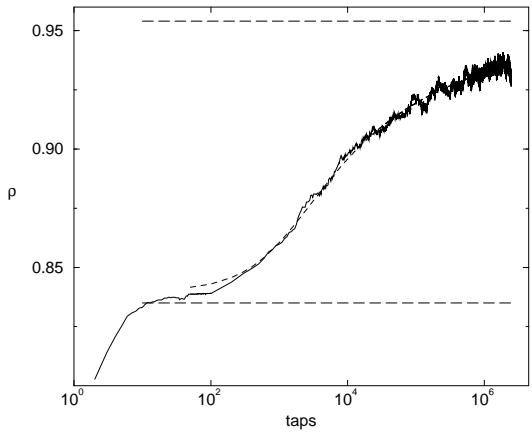}}
\caption{Compaction curve at connectivity $c=3$ for a system 
of $10^4$ spins (one spin is flipped at random per tap). 
The data stem from a single run with random 
initial conditions and the fit (dashed line) follows (\ref{loglaw}) 
with parameters $\rho_{\infty}=.971$, $\rho_0=.840$, $D=2.76$, 
and $\tau=1510$. The long-dashed line (top) indicates the approximate 
density $0.954$ at which the dynamical transition 
occurs, the long-dashed line (bottom) indicates the approximate 
density $0.835$ at which the fast dynamics stops, the 
{\it single-particle relaxation threshold}.}
\label{compact}
\end{figure}

An example of a single run of the system defined by (\ref{hdef}) is 
shown in Figure \ref{compact}. 
We can identify three regimes of this dynamics, first a very fast increase 
of the density up to a density $\rho_0$, then the slow compaction regime 
which takes the density up to $\rho_\infty$, and finally an asymptotic regime. 

In the first regime, which only 
lasts a few taps, all sites orient their spins in parallel with the local 
field acting on that site. This corresponds to a {\em fast} dynamics whereby 
\emph{single} particles \emph{locally} find the orientation maximizing the 
density leading to the density of $\rho_0$ 
\footnote{During the fast regime one may also model 
the mobility of the grains by a dynamics of the bonds subject to 
kinetic constraints; the resulting graph typically 
differs from those constructed according to 
(\ref{hdef}), however the compaction dynamics does not change 
qualitatively. }. 
We therefore term this the 
{\it single-particle relaxation threshold} (SPRT).  

This fast regime may be illustrated 
by considering a single site $i$ connected to $2k_i$ other sites 
and subject to the local field  
$h_i=1/2 \sum_{jk} C_{ijk} s_j s_k$. For random initial conditions, 
the values of $l_i=h_i s_i$ are binomially distributed with a probability 
of $C^{k_i}_{(k_i-l_i)/2} (1/2)^{k_i}$ if $k_i-l_i$ is even and zero if 
it is odd. If $l_i<0$ zero-temperature dynamics 
will flip this spin, turn $l_i$ to $-l_i$ and turn $(k_i\pm l_i)/2$ satisfied  
(dissatisfied) plaquettes connected to it into dissatisfied 
(satisfied) ones. This will cause the $l_j$ of $k_i\pm l_i$ neighbouring 
sites to decrease (increase) by 2. This dynamics stops when all sites 
have $l \geq 0$ giving $\rho_0=1/(3N) \sum_i l_i$. 

Neglecting loops in the random 
graph, the fast dynamics may be modelled by a simple population dynamics of 
$N$ elements, each with a Poisson 
distributed value of $k_i$ and a value of $l_i$ distributed according to 
the initial binomial distribution. At each step a randomly chosen 
element with negative $l_i$ has its $l_i$ inverted, and 
$k_i\pm l_i$ randomly chosen elements have their values of $l$ decreased 
(increased) by 2 until $l_i \geq 0 \ \forall i$. 
Running the population dynamics for $N=10^4$ at $c=3$ we obtain 
$\rho_0=0.835(1)$ which 
is shown as a dotted line in Figure \ref{compact}. Note that this 
density is found to be much higher than that of a typical 
'blocked' state 
with $l_i \geq 0 \ \forall i$ which is found to be 0.49
\cite{unpublished}(see also the discussion in \cite{barratkurchan}); 
despite the fact that these states
are exponentially dominant, the non-equilibrium, non-ergodic nature of the
fast dynamics is responsible for taking the system to an (atypical)
blocked state of higher density. 
At the end of the fast dynamics, there are no sites with 
more frustrated than unfrustrated plaquettes connected to them. 
The SPRT density thus appears as the density which is reached dynamically by putting 
each particle into its \emph{locally} optimal configuration, as has also been 
found in lattice-based models \cite{paj}
and simulations of sphere packings \cite{jpcm}, \cite{transunpub}, which 
incorporate both fast and slow dynamics.

The second phase of the dynamics consists of removing some of the remaining 
frustrated plaquettes and gives a logarithmically slow compaction 
\cite{nagel,paj,transunpub}
leading from density $\rho_0$ to $\rho_\infty$. 
The compaction curve may be fitted to the well-known logarithmic 
law \cite{nagel}
\begin{equation}
\label{loglaw}
\rho(t)=\rho_{\infty}-(\rho_{\infty}-\rho_0)/(1+1/D \, \ln(1+t/\tau)) \ ,
\end{equation} 
which may also be written in the simple form 
$1+t(\rho)/\tau = \exp{\{D \frac{\rho-\rho_0}{\rho_{\infty}-\rho} \}}\ , $
implying that the dynamics becomes slow (logarithmic) as soon as 
the density reaches $\rho_0$.  
With increasing density, free-energy barriers rise up causing 
the dynamics to slow down according to (\ref{loglaw}). The point where 
the height of these barriers scales with the system size marks a
{\em breaking of the ergodicity of the dynamics}, a break-up of the 
phase-space into 
a large number (scaling exponentially with the system size) of disconnected 
clusters, and a saturation of the 
compaction curve. For small driving amplitudes, we thus identify the 
asymptotic density (random close packing) with a {\em dynamical 
phase transition} \cite{cugliandolo,monassonfranzpar,aging}. 

To support this picture we give a simple approximation for the density 
$\rho_\infty$ at which the dynamical transition occurs. 
Using the replica-trick $\ln Z= \lim_{n \to 0} \partial_n Z^n$
\cite{MPV}
and 
standard manipulations we obtain for the average of the $n$-th power of the 
partition-function of the Hamiltonian (\ref{hdef}) averaged the ensemble of 
random graphs 
\begin{eqnarray}
\langle \langle &&Z^n \rangle \rangle = 
  \prod_{\vec{\sigma}} \int_0^1 dc(\vec{ \sigma}) \exp \left\{-N \left(
       \sum_{\vec{ \sigma}}c(\vec{ \sigma}) \ln(c(\vec{ \sigma})) \right. \right.  \\
       &&\left.\left.+ c/3
       - c/3 \sum_{\vec{ \omega},\vec{ \tau},\vec{ \sigma}}
       c(\vec{ \omega})c(\vec{ \tau})c(\vec{ \sigma}) \exp \{ \beta \sum_a 
       \omega^a \tau^a \sigma^a  \}
       \right)\right\} \ , \nonumber
\end{eqnarray}
where $c(\vec{\sigma})$ is an order parameter function defined on the 
domain of the $2^n$ vectors $\sigma^a=\pm1$. 
A simple variational ansatz \cite{BMW,hypersat,WZ} implementing one step of 
replica-symmetry breaking (RSB) 
\begin{equation}
c(\vec{ \sigma})=\prod_{b=1}^{n/m}  \left\{ \frac
       {\int dh^b G_{\Delta}(h^b) e^{\beta h^b \sum_{a=(b-1)m+1}^{bm}\sigma^a}}
       {\int dh^b G_{\Delta}(h^b)[2 \cosh(\beta h^b)]^m} \right\} \ ,
\end{equation}
where $G_\Delta(h)$ is a Gaussian with zero mean and variance $\Delta$, 
gives the free energy subject to the variational ansatz 
$f(\beta)=\mbox{extr}_{\Delta,m}f_1(\beta,\Delta,m)$ with  
\begin{eqnarray}
\beta&& f_1(\beta,\Delta,m)= \frac{\int Dz(\beta\sqrt{\Delta}z)
        [2\cosh(\beta\sqrt{\Delta}z)]^{m-1}
        \sinh(\beta\sqrt{\Delta}z) }
        {\int Dz [2\cosh(\beta\sqrt{\Delta}z)]^{m} } \nonumber \\
        &&-\frac{1-c}{m} \ln (\int Dz [2\cosh(\beta\sqrt{\Delta}z)]^{m}) 
        -c/(3m)\ln(\int \int \int Dz_1 Dz_2 Dz_3 \nonumber \\
        &&\left[8 \cosh(\beta\sqrt{\Delta}z_1) \cosh(\beta\sqrt{\Delta}z_2) 
        \cosh(\beta\sqrt{\Delta}z_3)\cosh(\beta) + 
         8 \sinh(\beta\sqrt{\Delta}z_1) \sinh(\beta\sqrt{\Delta}z_2) 
        \sinh(\beta\sqrt{\Delta}z_3)\sinh(\beta) \right]^m ) \ , \nonumber
\end{eqnarray}
where $D(z)$ denotes the Gaussian measure with zero mean and
variance one.
The dynamical transition occurs at a temperature 
where 
$\partial(\beta f(\beta,\Delta,m))/\partial m$ evaluated at $m=1$ 
develops a minimum at finite $\Delta$ \cite{monassonfranzpar}. The 
corresponding density is marked with a horizontal line in Figure 
\ref{compact} and agrees well with the asymptotic density reached by 
the tapping dynamics. 

The third regime of the compaction process is reached only asymptotically 
and consists of fluctuations around the random-close-packing
density. 

\begin{figure}
\myscalebox{\onefigure{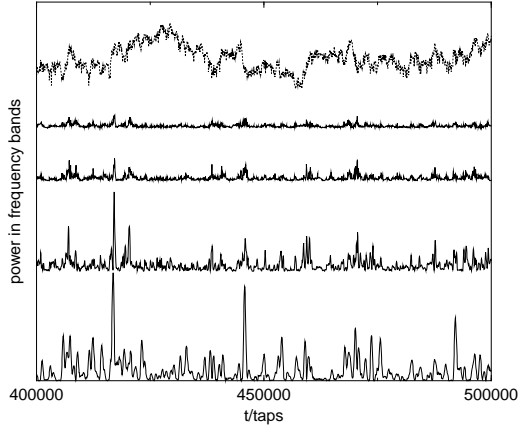}}
\caption{The Fourier-transform of the preceding 1024 taps is used to 
produce a power spectrum as a function of time. We plot the power in the 
first octave (frequency 1/(1024 taps) -2/(1024 taps)), second octave 
(frequency 2/(1024 taps) - 4/(1024 taps)), etc. up to the fourth octave 
(bottom to top). One finds that the fluctuations of the power in the 
different frequency bands are strongly correlated across many octaves, 
and correspond to sudden changes in the density (shown as dotted line).}
\label{freq}
\end{figure}

\begin{figure}
\myscalebox{\onefigure{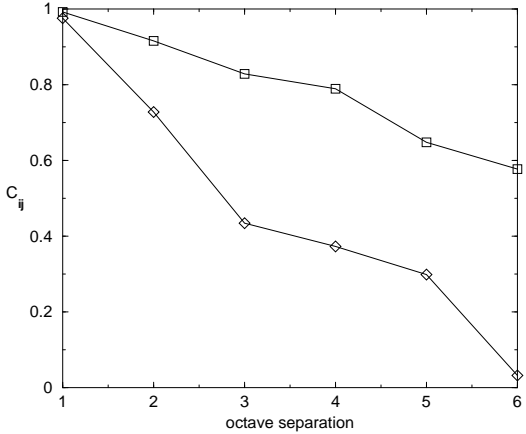}}
\caption{The rescaled covariances of the power fluctuations as 
are plotted as a function of the octave separation for both the 
ferromagnetic 3-spin model (squares) and the parking-lot model 
(diamonds). The definition of these quantities is provided in the 
text, a high value of the rescaled covariance indicates strong 
correlations of the power-fluctuations of two given frequency bands. 
}
\label{freqcorr}
\end{figure}

Figure \ref{compact} shows marked fluctuations around the logarithmic 
compaction law, which have been the subject of detailed experimental 
investigations \cite{ed}, where marked correlations between 
the Fourier components of different frequencies of these fluctuations 
were found. 
In order to compare the results of our model with experiment we follow 
\cite{ed} and use Fourier transforms to plot the power 
of the timeseries $\rho(t)$ in different frequency bands against time 
in Figure \ref{freq}. 
As in the experimental data, the fluctuation of the power 
in a given band shows marked 'bursts' after periods of calm. 
As a result of these bursts the fluctuations in the the power spectrum 
are correlated over a wide range of frequencies:  
In figure \ref{freqcorr} we have plotted, as in \cite{ed}, the average 
of the correlation matrix $C_{ij}$ as a function of octave separation $|i-j|$, 
where $C_{ij}:=M_{ij} \sqrt{ \frac{M_{ii} M_{jj}}{(M_{ii}-1) (M_{jj}-1)} }$ 
probes the non-Gaussian components of the correlation of the noise power in the 
$i$th and $j$th octaves. $M_{ij}$ is defined as the covariances of noise 
power fluctuations 
$\langle \delta O_i \delta O_j \rangle / 
\sqrt \langle (\delta O_i)^2\rangle \langle (\delta O_j)^2\rangle $ 
for $i \neq j$ and 
$M_{ii}:= \langle (\delta O_j)^2\rangle / \sum_{k \in i} \langle P_k \rangle^2$, 
where the average is over $5000$ time steps in the asymptotic regime, 
$\delta O_i$ are the power fluctuation around the average in the $i$th 
octave and $P_k, k \in i$ is the power in the $k$th frequency bin in octave $i$. 
Figure \ref{freqcorr} also shows the corresponding results for the parking-lot 
model \cite{plm}, where particles of a certain 
size are desorbed from and absorbed by a surface at random, subject to the constraint 
that none of the particles overlap. 

The experimental data analyzed in \cite{ed} 
shows a slow decay of the average $C_{ij}$ with octave separation (particularly in 
the bottom and top of the sample) with $C_{ij}$ decaying to about $0.6$ over $6$ 
decades. This behaviour is reproduced by the 3-spin model, but not by the 
parking-lot model. We argue that the 'burst' found in the time-series of the 
density is resposible for the correlation of noise power over a wide range of 
frequencies.  
In spin models with finite connectivity such bursts arise quite naturally 
due to \emph{cascades} of spin-flips. The crucial mechanism is that 
the flipping of a single spin alters the local fields acting on its 
neighbouring sites. 
The configuration of the spins on these sites may then 
no longer be locally stable, causing them to flip in turn. 
The first spin thus acts as a 'plug' releasing the neighbouring spins 
and setting off a cascade of successive spin-flips. 
A plug may also be composed of two or more  
sites, which need to have their spins flipped before neighbouring spins are 
released. 

Note that the cascade mechanism is entirely absent not only from 
generic fully connected models (where each spin interacts with 
all spins in the system, but with an interaction energy scaling as 
$1/\sqrt{N}$) but also from the parking-lot model: Clearly the  
creation and filling of such a gap by a particle does not cause the appearance 
of further gaps, in the way spins flips may trigger a cascade. For this 
reason, the bursts observed in experiment and in our model are absent in 
the parking-lot model as noted in \cite{ed}. 
In other, lattice-based models \cite{paj,tetris}  
of finitely connected granular media, however, cascades are also to be expected.  

To conclude, we have presented a \emph{finitely connected} 
spin model of vibrated granular matter, which
while reproducing the slow logarithmic relaxation associated
with compaction, has also thrown new light on the fast relaxation
mechanisms in granular dynamics.
Within this model we have also identified the asymptotic density 
reached with a dynamical phase transition, and interpreted the 
density-noise power fluctuations observed in recent experiments 
in terms of spin flip cascades.

\acknowledgments
AM is very grateful to the ICTP Trieste for providing an extremely pleasant
and stimulating environment where much of this work was carried
out. We thank A. Barrat, S. Franz, M. Leone, E. Nowak, 
F. Ricci-Tersenghi, P. Stadler, M. Weigt, and R. Zecchina
for illuminating discussions.

\end{document}